# Biot-Savart type magnetic field quantization via prime number theory, applications to symbolic dynamics

Y. Contoyiannis[1,2], P. Papadopoulos[1], N.L. Matiadou[1], S.M. Potirakis[1,3,*]

[1]Department of Electrical and Electronics Engineering, University of West Attica, Ancient Olive Grove Campus, 250 Thivon and P. Ralli, Athens GR12244, Greece
[2]Department of Physics, University of Athens, Athens GR15784, Greece
[3]Institute for Astronomy, Astrophysics, Space Applications and Remote Sensing, National Observatory of Athens, Metaxa and Vasileos Pavlou, Penteli, GR-15236 Athens, Greece
* Author to whom correspondence should be addressed: spoti@uniwa.gr

**Abstract:** In the present work we propose an algorithm based on the theory of prime numbers for the estimation of the magnetic field in a device of current carrying circular rings. Using the proposed algorithm, the magnetic field can be determined in a very good agreement with that resulting from an algorithm based on the Biot-Savart law. In addition, the prime-numbers-based algorithm gives quantized values of the magnetic field and reveals previously unknown behaviors such as special properties of the distribution of the waiting times at the quantized magnetic field values. Applications of the proposed prime-numbers-based algorithm to systems exhibiting symbolic dynamics is presented, proving its ability to provide a measure of the existence or not of dynamics in the system.

**Keywords**: Prime numbers; Biot-Savart magnetic field; Symbolic dynamics.

## 1. Introduction

The central forces of classical physics obey mathematical laws of the form $\sim \frac{1}{r^s}$, where $s = 2$ for the gravitational field (law of universal gravitation), electric field (Coulomb law), elementary magnetic field ($dB$) (Biot-Savart law). Given that in all three of these fundamental forces between bodies the sums of the forces appear in their calculations, i.e., quantities of the form $\sum_i \frac{1}{r_i^s}$, their convergence must be ensured. A more general expression of the harmonic series is the Riemann zeta function which is defined as $\zeta(s) = \sum_{i=1}^{\infty} \frac{1}{i^s}$, with $Re\{s\} > 1$ [1]. Thus, within the framework of a discretization of space with unit $l$, where $r_i \sim i \cdot l$ ($i = 1, 2, \dots$), the sum of the interactions can be treated as the Riemann zeta function.

One of the most famous unsolved issues in mathematics, which dates back to 1859, is the Riemann hypothesis [1,2], which asks where the zeros of the $\zeta(s)$ Riemann zeta function are located. This function is an analytic complex function. For complex numbers $s$ with real part $Re\{s\} > 1$, Riemann zeta function equals both an infinite sum over all integers, and an infinite product over the prime numbers. A natural number is called a *prime number* if it is $> 1$ and cannot be written as the product of two smaller natural numbers. Thus, by going one step further, we can "move" from the Riemann zeta function to the prime numbers through a theorem, known as the Euler product [3,4], according to which we can write:

$$\zeta(s) = \sum_{n=1}^{\infty} \frac{1}{n^s} = \prod_{p:prime} \frac{1}{1-p^{-s}} = \prod_{p:prime} \frac{p^s}{p^s-1}, \quad Re\{s\} > 1 \quad (1)$$

For a long time, the study of prime numbers, has been examined as the canonical example of pure mathematics, with no applications outside of mathematics. The concept of prime numbers is so important that it has been generalized in different ways in various branches of mathematics. Beyond the pure mathematics, the prime numbers are used in a series of various applications. Several public-key-cryptography algorithms, such as RSA and the Diffie-Hellman key exchange, are based on large prime numbers (2048-bit primes are common) [5]. Shor's algorithm can make any integer factor in a polynomial number of steps on a quantum computer [6]. Prime numbers are also used in pseudorandom number generators including linear congruential generators [7]. Beyond mathematics and computing, prime numbers have potential connections to quantum mechanics [8-11]. They have also been used in evolutionary biology to explain the life cycles of cicadas [12]. In this context, the present work is an application of prime numbers to devices for production of magnetic field based on the Biot-Savart law. The connection of zeta function [13] with the Biot-Savart theory has been presented in [14]. In the present work, we present the connection of Riemann zeta function with Biot-Savart theory and, furthermore, the connection with prime numbers for first time.

Out of the above-mentioned three central forces, the present work focuses to the magnetic field (Biot-Savart law). Specifically, it investigates to which extent can an algorithm based on prime numbers approach very close to the results of a physics-based algorithm computationally describing the magnetic field of a device of current carrying circular rings, which has extensively been studied in recent works [14,15]. The results of this investigation can be used as evidence of whether the classical central forces, in this case the Biot-Savart type magnetic field, are related (or not) to the theory of prime numbers. In addition, using such algorithms it is possible to perform simulations that can guide the real experiments, i.e., applications of the Biot-Savart laws in areas that they are currently unknown. In the present work it is shown that, with the appropriate approaches, one can reproduce the results of the Biot-Savart law through an algorithm based on the prime numbers. In fact, an improved version is produced, as concerns the values' convergence, which can lead to the quantization of the magnetic field. Applications of the prime-numbers-based algorithm (PNA) to systems that present symbolic dynamics can reveal quantitative characteristics of their dynamics, something that a physics-based algorithm (PA) isn't able to do.

## 2. Calculation of the magnetic field through the physics-based algorithm

In [14,15] a device of identical current carrying circular rings, shown in Figure 1, has been studied, through which electric currents of the same intensity $I$ flow, but their flow direction $(+I, -I)$ is randomly chosen, with equal probability.

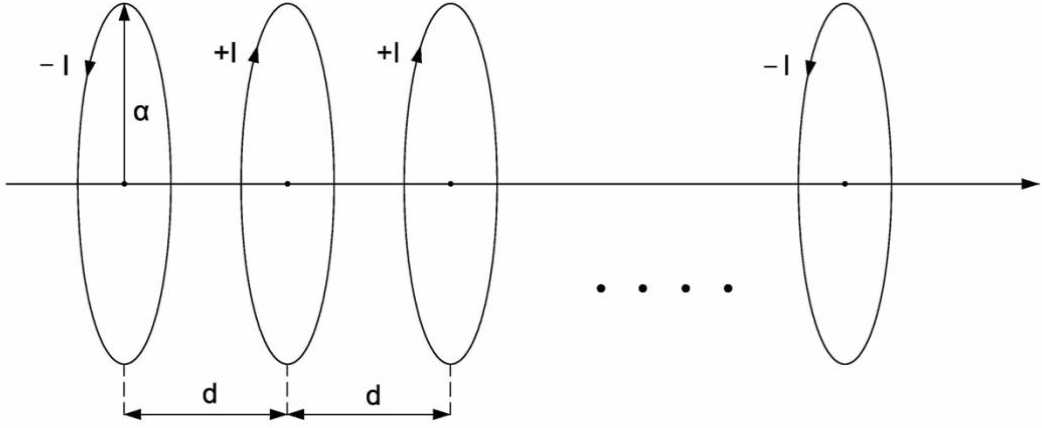

*Fig. 1. A device of identical rings through which currents of intensity I and random directions flow. The radius of the rings is $a$ and the distance of two consecutive rings is $d$.*

Assuming that the number of rings is $N$, then the magnetic field at the center of the $k-th$ ring ($k = 1,2, \ldots, N$) is of the following form [15]:

$$B_k = \frac{1}{2}\alpha^2 d^{-3} I \left\{ \sum_{n=0}^{k-1} \frac{I_{k-n}}{(n^2+c^2)^{3/2}} + \sum_{n=0}^{N-k-1} \frac{I_{n+1+k}}{[(n+1)^2+c^2]^{3/2}} \right\} \qquad (2)$$

where $\alpha$ is the radius of the rings, $d$ is the distance between two consecutive rings, $c = a/d$ and in each ring flows a current of magnitude $I$ and a random direction, $I_{k-n}, I_{n+1+k}$, described by a dichotomic variable taking the values $+1, -1$ with equal probability. The two sums appearing in the curly brackets of Eq. (2) can respectively be written as:

$$\sum_{n=0}^{k-1} \frac{I_{k-n}}{(n^2+c^2)^{3/2}} = \frac{I_k}{c^3} + \sum_{n=1}^{k-1} \frac{I_{k-n}}{(n^2+c^2)^{3/2}} \qquad (3)$$

and

$$\sum_{n=0}^{N-k-1} \frac{I_{n+1+k}}{[(n+1)^2+c^2]^{3/2}} = \sum_{m=1}^{N-k-1} \frac{I_{m+k}}{[m^2+c^2]^{3/2}} \qquad (4)$$

where $m = n + 1$.

Substituting Eqs. (3), (4) in Eq. (2) one gets:

$$B_k = \frac{1}{2}\alpha^2 d^{-3} I \left\{ \frac{I_k}{c^3} + \sum_{n=1}^{k-1} \frac{I_{k-n}}{(n^2+c^2)^{3/2}} + \sum_{m=1}^{N-k-1} \frac{I_{m+k}}{[m^2+c^2]^{3/2}} \right\} \qquad (5)$$

From Eq. (5), one can produce a physics-based algorithm (PA) based on the Biot-Savart law that calculates the magnetic field $B_k$ at all positions $k$ of the device's axis. As it has been shown in [14,15], the magnetic field inside the device is interesting when $c < 1$. Then, a stratification of the values of the magnetic field appears characterized by the existence of empty regions (cancelation) and zones where the fluctuations dominate.

In [15] the focus was on values of the ratio $c < 1$ and not on values $c \ll 1$, because the objective of that work was to present a realistic device for $c$ values that can realize the phenomenon of diffraction of the magnetic field. In the present work $c \ll 1$ values are also examined. In Fig. 2 we present the results of numerical experiments for the two cases ($c < 1$ and $c \ll 1$). In the whole article we will present the results in the positive half-plane of the diagram since, due to axial symmetry around $k$-axis, in the negative half-plane we get the same behavior.

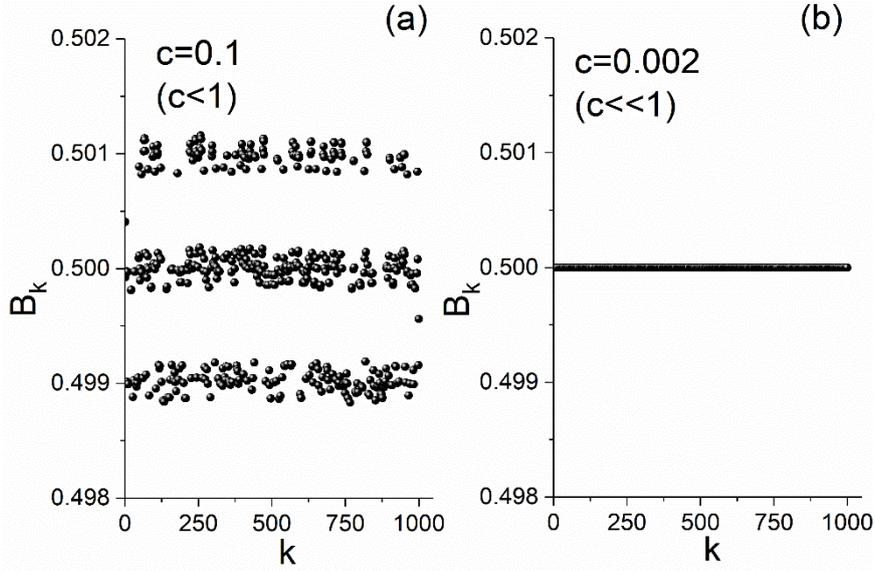

***Fig. 2*** *(a) For values $c < 1$ (here $c = 0.1$) three zones appear in the positive half-plane and respectively three symmetric zones in the negative half-plane (not shown here) within which the fluctuations of the magnetic field values is the key characteristic. (b) For values $c \ll 1$ (here $c = 0.002$) a fixed magnetic field (a single value) is produced at all positions of the device's axis (the symmetric picture appears in the negative half-level). The rest of the parameters' values for both simulation cases were: $I = 1$, $N = 1000$. The random current directions $I_{k-n}$, $I_{m+k}$ were described by a dichotomic variable taking the values $+1, -1$ with equal probabilities.*

It is obvious (Fig. 2a) that the PA has not converged for realistic values ($c < 1$), which is due to the fact that the alternating series, such as the ones included in Eq. (5), do not converge. However, for non-realistic values ($c \ll 1$), the algorithm converges to the magnetic field values $\pm 0.5$ (Fig. 2b). The device then behaves as a solenoid; the magnetic field inside the device is fixed at any position of its axis. In a realistic case ($c < 1$), the mechanism that produces this stratification of the values of the magnetic field has been explained in [15]. The explanation about the appearance of these zones is that something similar to the phenomenon of diffraction in optics, where some paths meet reinforcingly and some others paths meet catastrophically (cancelation), happens. The exact interpretation is presented in detail [15] in the framework of the magnetic field diffraction phenomenon. It would be important if one could quantize the magnetic field in the realistic $c$ values range ($c < 1$), for which the implementation of the specific device is feasible.

In Sec. 3 we investigate whether an appropriate approximation to Eq. (5), that would allow to use the convergence of harmonic series and consequently (see Eq. (1)) of the prime numbers, could meet the above mentioned objective. Of course, this way, the objective would have been achieved at the computer simulation level. However, that would also be enough to lead to useful software applications or to guide to useful conclusions about the real experiment.

## 3. Calculation of the magnetic field through the prime-numbers-based algorithm

Since our intention is to introduce the harmonic series to the magnetic field calculation, to lead to the prime numbers according to Eq. (1), we consider the case $c \ll 1$. Then, given the fact that $n, m \geq 1$ and using the condition $c \ll 1$, we can consider the approximations:

$$\sum_{n=1}^{k-1} \frac{I_{k-n}}{(n^2+c^2)^{3/2}} \approx \sum_{n=1}^{k-1} \frac{I_{k-n}}{n^3} \tag{6}$$

$$\sum_{m=1}^{N-k-1} \frac{I_{m+k}}{[m^2+c^2]^{3/2}} \approx \sum_{m=1}^{N-k-1} \frac{I_{m+k}}{m^3} \tag{7}$$

In Eqs. (6), (7) the directions of currents $I_{k-n}, I_{n+1+k}$ appear, which, as mentioned in Sec. 2, are described by a dichotomic variable taking the values $+1, -1$ with equal probability. This fact does not allow the calculation of the above sums, since these sums are randomly alternating sequences.

Let's assume that the random current directions hypothesis is suspended and all currents have, for example, positive direction, i.e., $I_{k-n} = +1$, $I_{n+1+k} = +1$, as happens for the solenoid case. In such a case one successively gets the following.

For high $k$ values, as $k \to \infty$, the sum in Eq. (6) is written:

$$\sum_{n=1}^{\infty} \frac{1}{n^3} = \zeta(3); \; Re\{\zeta(3)\} = 3 > 1 \tag{8}$$

For small $k$ values and as $N \to \infty$, the sum in Eq. (7) is written:

$$\sum_{m=1}^{\infty} \frac{1}{m^2} = \zeta(3); \; Re\{\zeta(3)\} = 3 > 1 \tag{9}$$

where $\zeta(3)$ is the Riemann zeta function $\zeta(s)$ for $s = 3$. Thus, one could introduce the prime numbers from the Eq. (1):

$$\zeta(3) = \prod_{p:prime \in \{1,\infty\}} \frac{1}{1-p^{-3}} = \prod_{p:prime \in \{1,\infty\}} \frac{p^3}{p^3-1} \tag{10}$$

As already mentioned, the effort here is to accomplish an appropriate approximation that allows to introduce the prime numbers to the estimation of the magnetic field of the device under study. The difficulty in the studied case, is to introduce the prime numbers but also to restore the random current directions hypothesis, by appropriately introducing the information of the randomness of the currents through the random alternations of the signs $+1, -1$.

The proposed solution is a prime-numbers-based algorithm (PNA) accomplished in two phases (the segment of the code which calculates the magnetic field through the proposed prime number approximation is presented in the Appendix):

In the first phase (Phase A), the random direction quantities $I_{k-n}, I_{m+k}$ are not taken into account. Then, the series of Eqs. (6), (7) are harmonic series as the series of Eqs. (8), (9), which converge and then the sums are Riemann zeta functions and due to Eq. (10) the prime numbers are introduced.

In the second phase (Phase B), an iterative procedure is proposed, comprising two loops that randomly produce the signs of $I_{k-n}, I_{k+m}$.

In Fig. 3 we present the results from numerical experiments through the above described PNA for the case $c = 0.1$ ($c < 1$) and the case $c = 0.002$ ($c \ll 1$).

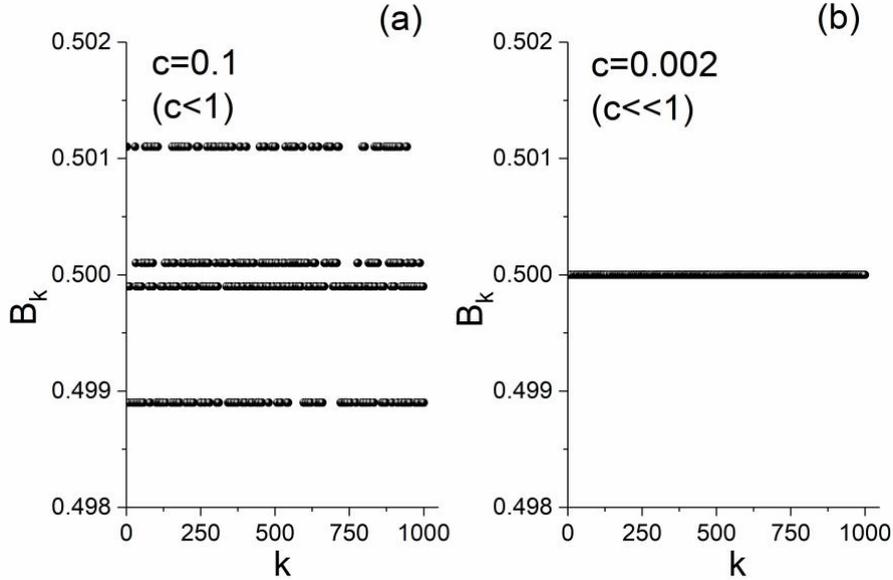

*Fig. 3 (a) For the value $c = 0.1$ ($c < 1$) a central zone around the value $B_k = 0.500$ appears where a fine structure phenomenon of fixed value levels exists. Two more fixed value levels appear, one above and one below the central zone. (b) For value $c = 0.002$ ($c \ll 1$) a fixed level of a single value results at $B_k = 0.500$. The rest of the parameters' values for both simulation cases were: $I = 1$, $N = 1000$. The random current directions $I_{k-n}, I_{m+k}$ were described by a dichotomic variable taking the values $+1, -1$ with equal probabilities. Note that the axis symmetry mentioned in Fig. 2, also appears here, although only the positive half-plane results are shown.*

The validity of the approximation to the calculation of the magnetic field through the proposed

PNA approach would be verified if PNA could lead to exactly the same results as the PA for the same value of $c = 0.002$, for which the PA has converged. This indeed happens, as evident by comparing Fig. 2b with Fig. 3b. Furthermore, due to the absolute coincidence of the results produced by the two algorithms for $c \ll 1$, one could, as an approximation to the convergence observed in the case of PA, use the PNA to roughly describe the realistic case of $c < 1$. This would be a reasonable approach if the average values of the zones of the PA results for $c < 1$ were very close to the values of the quantized fixed levels of the PNA results for the same, $c < 1$, $c$ value. By comparing Fig. 2a with Fig. 3a, which present the numerical experiment results for $c = 0.1$ by PA and PNA, respectively, this is indeed verified.

## 4. Comments on the results obtained by the two examined algorithms

From Figs 2b and 3b it results that for $c \ll 1$ the two algorithms actually converge to the same fixed magnetic field values. This result means that for $c \ll 1$ the device, in which we introduce currents of equal intensity but of random direction, behaves like a solenoid presenting a fixed value of magnetic field $B_k = \pm 0.5$ at any position of the device's axis, for the values of the parameters which have been used for the two numerical experiments. Therefore, the PNA gives exactly the same outcome with the PA in the unrealistic cases of devices with $c \ll 1$.

For realistic cases of devices with $c < 1$ the following behavior is observed:

(a) The PA ends up to zones of magnetic field value's fluctuations, while the PNA ends up to specific levels of fixed magnetic field value in place of the corresponding PA zones, i.e., the PNA "quantizes" the magnetic field.

(b) At the central zone, the PNA gives rise to a phenomenon known from the physics of quantized levels as a "fine structure", by dividing the central level into two levels in very close positions. A detailed examination of the PA results could be said to present a similar behavior for the central zone but the strong fluctuations do not allow for a clear picture.

(c) The fixed values resulting from the PNA are very close to the average values of the fluctuations of each corresponding zone obtained by means of the PA. Specifically, these mean values coincide, up to the third decimal place, with the corresponding constant PNA values: $B_k = 0.501$ for the upper zone, $B_k = 0.499$ for the lower zone, while for the central zone in the PA case the average value is $B_k = 0.500$ and in the PNA case a mean level of the revealed fine structure is also at $B_k = 0.500$. These results are considered a very satisfactory approach to physical reality for realistic devices.

(d) The probabilities with which the PA calculated magnetic field values are distributed to the three fluctuation zones are 24%, 45.5% and 30.5% for the upper, central and lower zone, respectively. Interestingly, for the case of PNA the corresponding probabilities, if we consider that both the revealed fine structure levels are accounted as a "central zone", are 23%, 46% and 31%. These results are almost the same for both algorithms, confirming the fact that the prime numbers approximation is very close to the real results. The above results have been recorded for $N = 1000$ rings. If ones approaches the asymptotic limit (for example $N = 30000$) then both algorithms converge in the probability ratio 1:2:1 for upper, central and lower zones, respectively.

## 5. Application of PNA to symbolic dynamics sequences

Symbolic dynamics refers to a description of complex systems, according to which a complex system is considered as an information generator producing messages constituted of a discrete set of symbols defined by partitioning the full continuous phase space into a finite number of cells, thus implementing a coarse graining strategy. The simplest possible coarse graining corresponds to the assignment of just two symbols "0" and "1", or "−1" and "+1", etc., to the original time series, depending on whether it is above or below a specific threshold (binary partition). On the other hand, some physical or numerical systems are already described in terms of discrete states, e.g., spin systems or DNA sequences, so these inherently fit to the symbolic dynamics description.

In the following the idea of replacing the random current directions hypothesis with dynamic current directions hypothesis, i.e., when current directions' sequence is determined by a physical or numerical system symbolic dynamics, is investigated. Specifically, the "introduction" of a two symbols symbolic time series into the device of current carrying circular rings under study, is achieved by corresponding the chronological order of the symbols of the symbolic time series to the positions $k$ of the rings, while the symbol of each specific time point is used to determine the direction of the ring's current at the corresponding position. Then, the magnetic field values of the device are calculated by PNA and the distribution of the lengths $L$ of the segments of successive fixed values of the magnetic field (i.e., "waiting times" at a specific fixed magnetic field value) are studied. The objective of the described investigation is to find out whether the symbolic dynamics of the currents' directions sequence are reflected on the magnetic field values as estimated by the PNA, by quantifying how far or close is the dynamics of the system to randomness.

The information provided by the waiting times' distribution is important for time series analysis, including the analysis of symbolic dynamics time series. In the case of the simplest possible coarse graining of only two symbols, the absence of fluctuations renders the calculation of waiting times a simple process of counting consecutive values above or below a specific threshold. It is known [16,17] that if the distribution of quantities such as the waiting times, produced from a time series, is exponential then randomness dominates in the time series. In the rest of this paper, the lengths of the fixed value segments of the magnetic field, as these are produced by the PNA, are considered analogous to the waiting times in a time series. Exponential distributions of the lengths of waiting times means that the long lengths are cut. Thus, in such cases the long-range correlations and the dynamics produced by them are absent. The quantitative evaluation of how close is the system to randomness can be made by means of the value of the negative exponent of the exponential distribution. The more negative the exponent is, the narrower is the (short) lengths range included in the distribution, the closer the system is to randomness. At the opposite end to the absence of dynamics, the "full dynamics" case exists, which allows the presence of all scales of lengths $L$, from the very short, up to lengths equal to the size of the system. The distribution of these lengths is mathematically expressed by power-laws. Thus, between the two extreme behaviors, that is, exponential and power-law, all real systems' dynamic behaviors, according to the criterion of the distribution of waiting times, can be found.

At this point we would like to explain the reason why PNA is permitting the aforementioned investigation, which would be impossible if one would use PA instead. The PNA allows for a more in-depth analysis for the reason that leads to quantized magnetic field values giving rise to a fine description by calculating the waiting times at specific fixed values or even groups of fixed values, which permits for smaller scales of waiting times to be formed. This is especially important in the cases that one examines whether the symbolic dynamics come from a critical system. In order to decide on the criticality, the coarse extraction of power laws is not enough; it is absolutely necessary to show that these power laws extend to fine structure too, i.e., on smaller scales. In the case of symbolic dynamics, one has no other option to find such scales than with PNA, which by "quantizing" the values of the magnetic field indicates where to look. Thus, we consider that the detection of dynamics, especially of the critical dynamics, is possible by the proposed approach. The PA is not allowing for such an analysis of coarse-grained time series because the resulting magnetic field values fluctuate and, thus, such segments, within which the successive magnetic field values are fixed, do not appear.

In the following, two examples of application of the PNA are presented: (a) to a well-known numerical system such as the Ising model of phase transition phenomena (Sec. 5.1) and (b) to a biological system such as DNA (Sec. 5.2). In the presented examples the consistency of PNA with the results expected from the coarse description is shown, while in case of critical dynamics we proceed to the confirmation that includes the aforementioned fine descriptions.

**5.1 The 2D-Ising model case**

For a Z(N) spin system, spin variables are defined as: $s(a_i) = e^{i2\pi a_i/N}$ (lattice vertices $i = 1 \ldots i_{max}$) with $a_i = 0,1,2,3 \ldots N-1$. Specifically, for $N = 2$ and for 2 dimensions we consider the 2D-Ising model. An effective algorithm which produces configurations for the 2D-Ising model is the Metropolis algorithm. According to this algorithm the configurations at constant temperatures are selected with Boltzmann statistical weights $e^{-\beta H}$, where $H$, the Hamiltonian of the spin system with nearest neighbors' interactions, can be written as:

$$H = -\sum_{<i,j>} J_{ij} s_i s_j \tag{11}$$

It is known [18] that this model undergoes a second-order phase transition when the temperature drops below a critical value. Thus, for a $100^2$ lattice the critical (or pseudocritical for finite size lattices) temperature has been found to be $T_c = 2.308$ ($J_{ij} = 1$). The sweep of the whole lattice represents the algorithmic time unit. The possible values that the spin takes in the model are $\pm 1$. One can produce a time series of symbolic dynamics [19,20] with two symbols, "+1", "−1", by randomly selecting a position of the lattice and monitoring, vs. the algorithmic time, the evolution of the spin at the specific position. It is of particular interest when the production of such a time series takes place at critical temperatures. It is known that at critical temperatures the power laws dominate the size distributions, such as temporal and spatial lengths and especially quantities that have the character of waiting times (laminar lengths) [21].

In the following, the above described 2D-Ising sourced symbolic time series is "introduced" into the device of current carrying circular rings under study as described in Sec. 5. The magnetic field values are estimated by using the PNA and the resultant distributions of the fixed value segments are compared to the corresponding distributions in the case that the directions of ring's currents are random, determined by a dichotomic variable taking the values $+1, -1$ with equal probabilities.

Figure 4 presents the results of the PNA application when the currents' directions are the randomly selected (Fig. 4a) and when the currents' directions follow the aforementioned spin symbolic dynamics (Fig. 4b). The waiting times refer to lengths of successive positive or negative magnetic field values, without distinguishing any internal structures. It's a coarse graining description of two symbols "positive magnetic field value" and "negative magnetic field value", i.e., by using as threshold zero magnetic field. The simulation parameters' values were: $a = 1$, $d = 10$, $c = 0.1, I = 1$. Since large statistics are necessary for the production of distributions, $N = 30000$ was used, for which, according to Sec. 4, there has been convergence of the probabilities of the quantized levels of the values of the magnetic field.

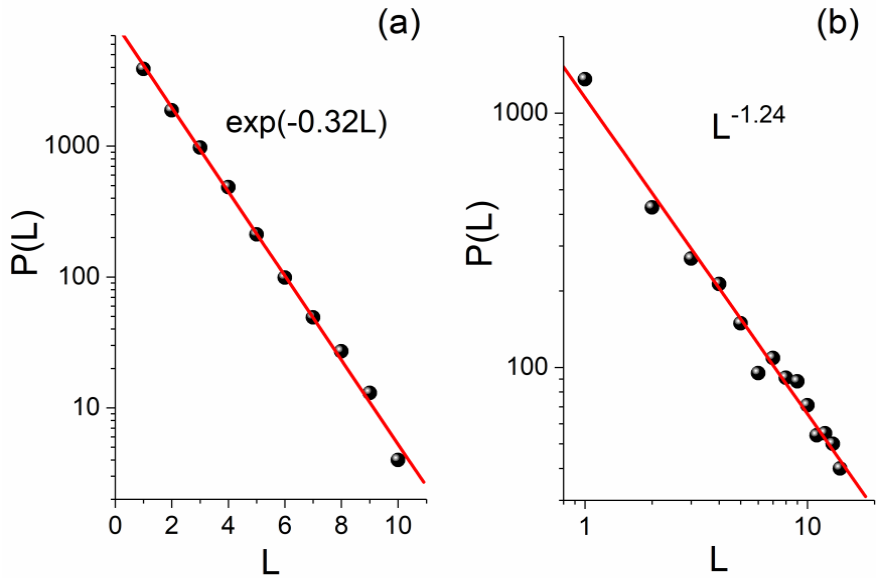

***Fig.4*** *The distributions of waiting times for the four discrete positive values of the magnetic field (shown in Fig. 3a): (a) for random currents' directions, resulting to an exponential decay with exponent –0.32. (b) for currents' directions determined by the 2D-Ising symbolic dynamics in spin lattice site (15,88), a power-law distribution $P(L) \sim L^{-1.24}$ for the first 14 points is shown.*

From Figs. 4a and 4b it can be confirmed that the symbolic dynamics of the currents' directions sequence are indeed reflected on the magnetic field values as estimated by the PNA. Therefore, the PNA is consistent with the results expected from the considered coarse descriptions, which means that it preserves the two following basic behaviors of dynamic systems, that is:

(a) the exponential distribution of waiting times in "time series" that randomness dominates, and

(b) the power-law distribution of waiting times in "time series" that are produced from critical states (critical points) of natural systems.

As already mentioned, the above presented example refers to a coarse graining description of the magnetic field, where the two symbols were considered by using zero magnetic field as a threshold. Figure 5 shows the distributions of waiting times for a more detailed analysis, by examining two more thresholding approaches: (a) one of the two symbols corresponds to the lower positive level of the quantized magnetic field values (see Fig. 3a) and the other symbol to all other levels (positive and negative), (b) one of the two symbols corresponds to the central levels (see Fig. 3a) of the revealed fine structure and the other symbol to all of other levels (positive and negative).

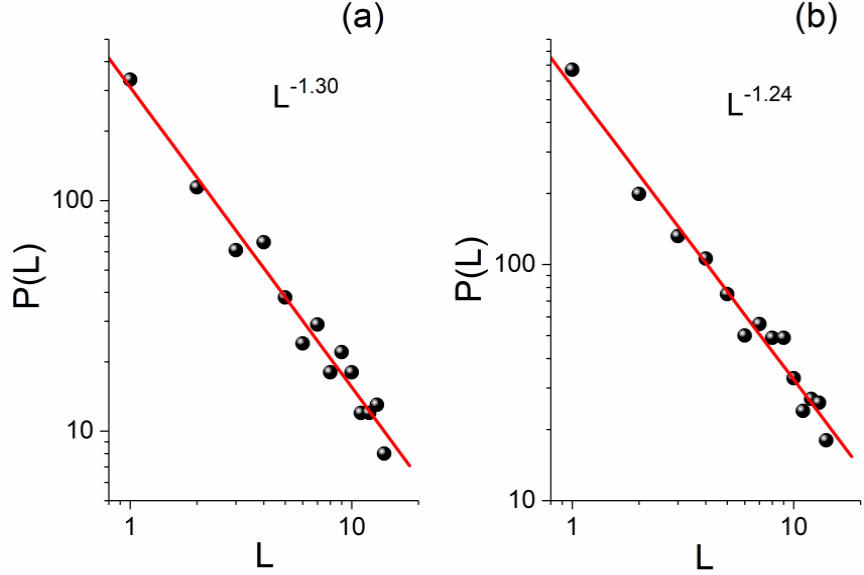

*Fig. 5*. The distributions of waiting times for two different thresholding approaches than that of Fig. 4. (a) One of the two symbols corresponds to the lower positive level of the quantized magnetic field values (see Fig. 3a, $B_k$ value 0.498899) and the other symbol to all other levels (positive and negative). (b) One of the two symbols corresponds to the two central levels (see Fig. 3a, $B_k$ values 0.499899 and 0.500101) of the revealed fine structure and the other symbol to all of other levels (positive and negative).

We chose to show the two extreme distributions of waiting times, i.e., exponential and power-law, to highlight the ability of the PNA to distinguish them. As mentioned in Sec. 5, the method can be autonomous and used to provide a quantitative indication of how close or far the dynamic of a system is from each end, i.e., randomness, on the one hand, and extended dynamics on all scales as it appears at the critical point on the other hand.

In order to demonstrate the ability of the PNA to respond to changes in the time series that determines the current directions' sequence we present Fig. 6. Specifically, Fig. 6a shows the deviation from the power-law of the Fig. 4b if the 2D-Ising time series is produced for a temperature higher than the preudocritical ($T = 3.2 > T_c$), whereas Fig. 6b shows the distribution of waiting times if we impose in the data which gave the power-law of the Fig. 4b a form of shuffling (surrogate type). In the latter case, an exponential distribution results with an exponent close to the exponent of the randomness of the Fig. 4a.

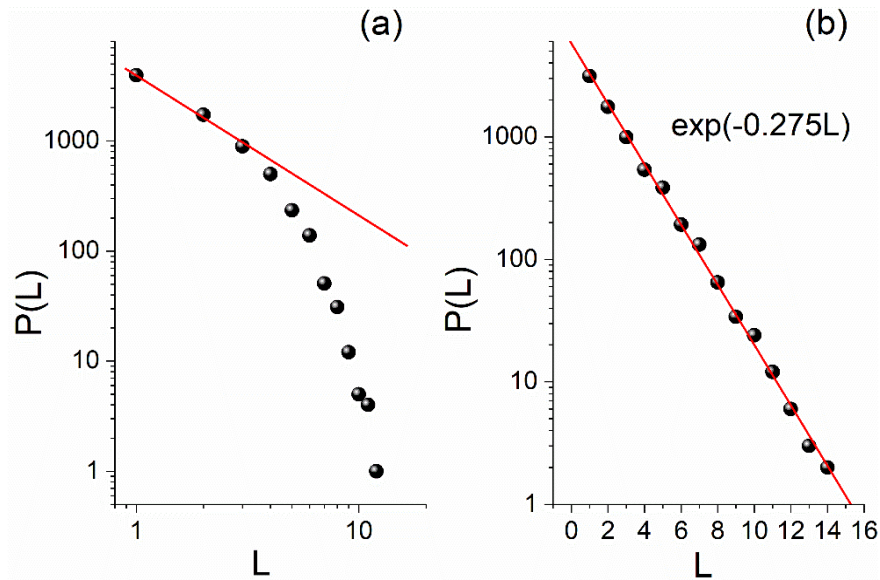

*Fig. 6* (a) *The deviation from the power-law of Fig. 4b for the waiting times distribution when the temperature in 2D-Ising model has not its critical value. (b) After a surrogate type procedure in the data which give the power law of Fig. 4b the power-law is destroyed and an exponential distribution for the waiting times appears.*

**5.2 The DNA case**

The example presented in this section refers to a human (Homo sapiens) gene. The DNA is a sequence of 4 bases, Adenine, Guanine, Cytosine and Thymine denoted by the letters A, G, C, T, respectively. Bases A, G belong to the category of purines and bases C, T to the category of pyrimidines. Thus, we could express the DNA sequence of the GAPDH (Glyceraldehyde-3-Phosphate Dehydrogenase) gene of the Homo sapiens, as a sequence of purines and pyrimidines, that is as a "time series" of symbolic dynamics with the symbols $+1, -1$. After turning the gene into a symbolic dynamic "time series", the PNA was applied as presented in Sec. 5. The results are shown in Fig. 7.

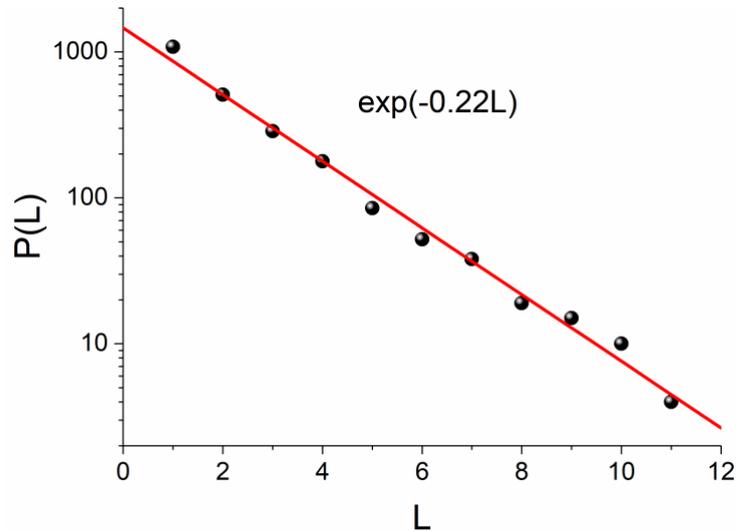

*Fig. 7 The distribution of waiting times produced via the PNA when the currents' direction is determined by the symbolic dynamics of the GAPDH (Glyceraldehyde-3-Phosphate Dehydrogenase) gene of Homo sapiens.*

As shown in Fig. 7 the exponent value of the exponential distribution is higher than the corresponding exponent value of a random sequence. This means that longer lengths survive. So, in contrast to a random sequence, human gene presents some kind of structure. The correlations that are responsible for this structure must have organized the coding part of the gene. Finding quantitative relationships between a large number of genes through this PNA methodology is a challenging task for a future study.

In this example we saw that a pure mathematical theory such as prime numbers, combined with a device of physics such as a device of current carrying circular rings, is able of extracting biological information from a biological structure such as DNA.

## 6. Conclusions

According to the results presented in this work, it is clear that the approximation of the proposed prime-numbers-based algorithm (PNA) completely agrees with the physical algorithm that calculates the magnetic field inside a ring device for unrealistic values $c \ll 1$ and that it closely follows the results for realistic cases where $c < 1$. The quantization of the magnetic field values offered by PNA can give useful simulations that guide real experiments in the region of Biot-Savart magnetic forces, such as the examined device of current carrying circular rings. An extension to systems exhibiting symbolic dynamics of two symbols is presented. By comparing the PNA obtained results when the current directions of the device are defined by a symbolic dynamics time series with the results for the case of random current directions' sequence, one can have a measure of the existence or not of dynamics in the system which produces the specific time series. PNA could be seen as a kind of "transformation" which can transform any two symbols coarse-grained symbolic dynamics time series to a finer structure in the form of the quantized magnetic field of the ring device under study.

We consider that the suggested approach offers a new perspective in the study of complex systems that could offer a unified way of studying diverse complex systems, which is something that remains to be explored in depth in the future.

But beyond the applications, the introduction of the theory of prime numbers in the study of natural phenomena is in itself an important fact that has conceptual extensions. As an example of an interesting future study, we refer to the main problem in number theory is to understand the distribution of prime numbers. Many deep problems in analytic number theory can be expressed in terms of the prime counting function $\pi(x)$ (defined as the number of primes less than or equal to $x$), for example the Riemann hypothesis. One of our future research goals is the investigation for the presence of $\pi(x)$ in the study of natural phenomena.

**Authors' Contributions**

All authors contributed equally to this work.

**Data Availability**

The data that support the findings of this study are available upon request to the authors.

**Acknowledgement**

The authors would like to thank Prof. E. Kosmidis for providing the DNA data.

**Appendix**

FORTRAN code for the calculation of the magnetic field through the proposed prime number approximation, the prime-numbers-based algorithm (PNA).

1. Code for the production of the random current directions $I_{k-n}$, $I_{m+k}$, described by a dichotomic variable taking the values $+1, -1$ with equal probabilities.

```
*********Random currents production***************************
do L=1,NuRings
 call random(rnd)
 if(rnd.le.0.5) then
I(L)=1
 else
I(L)=-1
 endif
enddo
***************************************************************
```

2. Code for the calculation of the series of Eqs. (8), (9) during Phase A of PNA. The introduction of prime numbers has been done in the form of vector elements: prime(i) for i=1 to 158 (for the first 1000 integer numbers).

```
*********** Phase A: Prime numbers*****************************
piA=1
 do i=1,158
piA=piA*((prime(i)**3)/(prime(i)**3 -1))
enddo
piB=1
 do i=1,158
piB=piB*((prime(i)**3)/(prime(i)**3 -1)
enddo
****************************************************************
```

2. Code for the introduction of currents randomness during Phase B of PNA. **NuRings** is the number of Rings, **I** is the current of k ring, **Beta** is the magnetic field, **coef** is the quantity $\frac{1}{2}\alpha^2 d^{-3} I$.

```
********** Phase B: Introduction of currents randomness*********
do k=1,NuRings
do n=1, k-1
piA=I(k-n)*piA
enddo
 do m=1, NuRings-k-1
piB=I(m+k)*piB
        enddo
 Beta(k)=coef*(I(k)*(c**(-3))+piA+piB)
enddo
****************************************************************
```

# References

[1] Bombieri, E., The Riemann Hypothesis – official problem description, Clay Mathematics Institute,

Retrieved (8/8/ 2014).

[2] Borwein, P., Choi, St., Rooney, B. and Weirathmueller, A., The Riemann Hy-pothesis: A Resource for the Afficionado and Virtuoso Alike, Springer, New York, 10 (2008).

[3] Patterson, S.J., Cambridge Studies in Advanced Mathematics 14, Cambridge University Press, Cambridge, 1 (1988).

[4] Derbyshire J., Prime Obsession: Bernhard Riemann and the Greatest Unsolved Problem in Mathematics, Washington, DC: Joseph Henry Press, Chapter 7, (2003).

[5] Chirgwin, R., Crypto needs more transparency, researchers warn, The Register, (2016).

[6] Rieffel, E., Polak W., Quantum Computing: A Gentle Introduction, MIT Press, 163 (2011).

[7] Knuth D. E., 3.2.1 The linear congruential model, The Art of Computer Programming, Vol. 2: Seminumerical algorithms (3rd ed.), Addison-Wesley, 10 (1998).

[8] Peterson I., The Return of Zeta, MAA Online, (1999).

[9] Hayes B., Computing science: The spectrum of Riemannium, American Scientist, 91 (4), 296 (2003).

[10] Bengtsson I., Życzkowski K., Geometry of quantum states : an introduction to quantum entanglement, Cambridge: CambridgeUniversityPress, (2017).

[11] Zhu H., SIC POVMs and Clifford groups in prime dimensions, Journal of Physics A: Mathematical and Theoretical, 43(30):305305, (2010).

[12] Goles E., Schulz O., Markus M., Prime number selection of cycles in a predator-prey model, Complexity, 6 (4): 33 (2001).

[13] Abramowitz M., Stegun I.A., Handbook of Mathematical Functions, Dover, Washington, (1970).

[14] Anagiannis V., Contoyiannis Y., Diakonos F., Magnetic field fluctuations in an array of randomly directed circular currents, The European Physical Journal B, (2013).

[15] Contoyiannis Y., Potirakis S.M., Papadopoulos P., Kampitakis M., Diffraction-like stratified magnetic field in a device of circular rings, Journal of Applied Physics, (2021).

[16] Contoyiannis. Y., Kapiris P., Eftaxias K., Monitoring of a preseismic phase from its electromagnetic precursors, Phys. Rev. E 71, 1(2005).

[17] Contoyiannis Y., Diakonos F., Papaefthimiou C., Theophilidis G., Criticality in the relaxation phase of the spontaneous contracting atria isolated from the heart of the frog (Rana ridibunda), Phys. Rev. Lett. 93, 098101 (2004).

[18] Huang K., Statistical Mechanics, 2nd edn New York: Wiley, (1987).

[19] Kitchens B., Symbolic dynamics. One-sided, two-sided and countable state Markov shifts, Springer-Verlag, Berlin, (1998).

[20] Lind D., Marcus B., An introduction to symbolic dynamics and coding, Cambridge University Press, (1995).

[21] Contoyiannis Y., Diakonos F., Malakis A., Intermittent dynamics of critical fluctuations, Phys.Rev. Lett. 89, 35701(2002).